\documentclass[twocolumn,tighten,times,twocolappendix]{aastex631_modified}

\usepackage[fontsize=9.6pt]{fontsize}

\usepackage{amssymb,amsmath}
\usepackage{bm}
\usepackage{multirow}
\usepackage{nicefrac}
\shortauthors{Schad et al.}
\newcommand{\uat}[2]{\href{http://astrothesaurus.org/uat/#2}{#1 (#2)}}
\usepackage{array}
\journalinfo{Accepted for publication in ApJ | Current Version Date: \today}
\usepackage{booktabs}

\usepackage{setspace}
\usepackage{ragged2e}

\graphicspath{{./}}

\DeclareRobustCommand{\okina}{%
  \raisebox{\dimexpr\fontcharht\font`f-\height}{%
    \scalebox{0.8}{`}%
  }%
}

\usepackage{etoolbox}

\makeatletter

\patchcmd{\NAT@citex}
  {\@citea\NAT@hyper@{%
     \NAT@nmfmt{\NAT@nm}%
     \hyper@natlinkbreak{\NAT@aysep\NAT@spacechar}{\@citeb\@extra@b@citeb}%
     \NAT@date}}
  {\@citea\NAT@nmfmt{\NAT@nm}%
   \NAT@aysep\NAT@spacechar\NAT@hyper@{\NAT@date}}{}{}

\patchcmd{\NAT@citex}
  {\@citea\NAT@hyper@{%
     \NAT@nmfmt{\NAT@nm}%
     \hyper@natlinkbreak{\NAT@spacechar\NAT@@open\if*#1*\else#1\NAT@spacechar\fi}%
       {\@citeb\@extra@b@citeb}%
     \NAT@date}}
  {\@citea\NAT@nmfmt{\NAT@nm}%
   \NAT@spacechar\NAT@@open\if*#1*\else#1\NAT@spacechar\fi\NAT@hyper@{\NAT@date}}
  {}{}

\makeatother

\newlength{\bibitemsep}
\newlength{\bibparskip}
\let\oldthebibliography\thebibliography
\renewcommand\thebibliography[1]{%
  \oldthebibliography{#1}%
  \setlength{\parskip}{\bibitemsep}%
  \setlength{\itemsep}{\bibparskip}%
}

\begin{document}
\title{\Large{Coronagraphic observations of \ion{Si}{10} 1430 nm acquired by DKIST/Cryo-NIRSP \\ with methods for telluric absorption correction}}
\setstretch{0.95}

\correspondingauthor{Thomas A. Schad}
\email{tschad@nso.edu}
\author[0000-0002-7451-9804]{Thomas A. Schad}
\affiliation{National Solar Observatory, 22 \okina\={O}hi\okina a K\={u} Street, Pukalani, HI 96768, USA}
\author[0000-0002-7978-368X]{Andre Fehlmann}
\affiliation{National Solar Observatory, 22 \okina\={O}hi\okina a K\={u} Street, Pukalani, HI 96768, USA}
\author[0000-0002-6003-4646]{Gabriel I. Dima}
\affil{Cooperative Institute for Research in Environmental Sciences, CU Boulder, CO 80309, USA}
\affil{NOAA National Centers for Environmental Information, DSRC, 325 Broadway, Boulder, CO 80305, USA}
\author[0000-0003-1361-9104]{Jeffrey R. Kuhn}
\affiliation{University of Hawai\okina i, Institute for Astronomy, 34 \okina\={O}hi\okina a K\={u} Street, Pukalani, HI 96768, USA}
\author{Isabelle F. Scholl}
\affiliation{National Solar Observatory, 22 \okina\={O}hi\okina a K\={u} Street, Pukalani, HI 96768, USA}
\author[0000-0002-3215-7155]{David Harrington} 
\affiliation{National Solar Observatory, 22 \okina\={O}hi\okina a K\={u} Street, Pukalani, HI 96768, USA}
\author[0000-0002-7213-9787]{Thomas Rimmele}
\affiliation{National Solar Observatory, 3665 Discovery Drive, Boulder, CO, 80303, USA}
\author[0000-0003-3147-8026]{Alexandra Tritschler} 
\affiliation{National Solar Observatory, 3665 Discovery Drive, Boulder, CO, 80303, USA}
\author[0000-0002-3491-1983]{Alin R. Paraschiv} 
\affiliation{National Solar Observatory, 3665 Discovery Drive, Boulder, CO, 80303, USA}

\begin{abstract}
\noindent We report commissioning observations of the \ion{Si}{10} 1430 nm solar coronal line observed coronagraphically with the \textit{Cryogenic Near-Infrared Spectropolarimeter} (Cryo-NIRSP) at the National Science Foundation's \textit{Daniel K. Inouye Solar Telescope} (DKIST).  These are the first known spatially resolved observations of this spectral line, which has strong potential as a coronal magnetic field diagnostic.  The observations target a complex active region located on the solar northeast limb on 4 March 2022.  We present a first analysis of this data, which extracts the spectral line properties through a careful treatment of the variable atmospheric transmission that is known to impact this spectral window.  Rastered images are created and compared with EUV observations from the SDO/AIA instrument.  A method for estimating the electron density from the \ion{Si}{10} observations is then demonstrated that makes use of the forbidden line's density-sensitive emissivity and an emission-measure analysis of the SDO/AIA bandpass observations.  In addition, we derive an effective temperature and non-thermal line width across the region.  This study informs the calibration approaches required for more routine observations of this promising diagnostic line.
\end{abstract}
\keywords{\uat{Solar corona}{1483}; \uat{Solar coronal lines}{2038}; \uat{Solar E corona}{1990}}

\section{Introduction} \label{sec:intro}

\begin{figure*}
\centering
\includegraphics[width=0.85\textwidth,clip=]{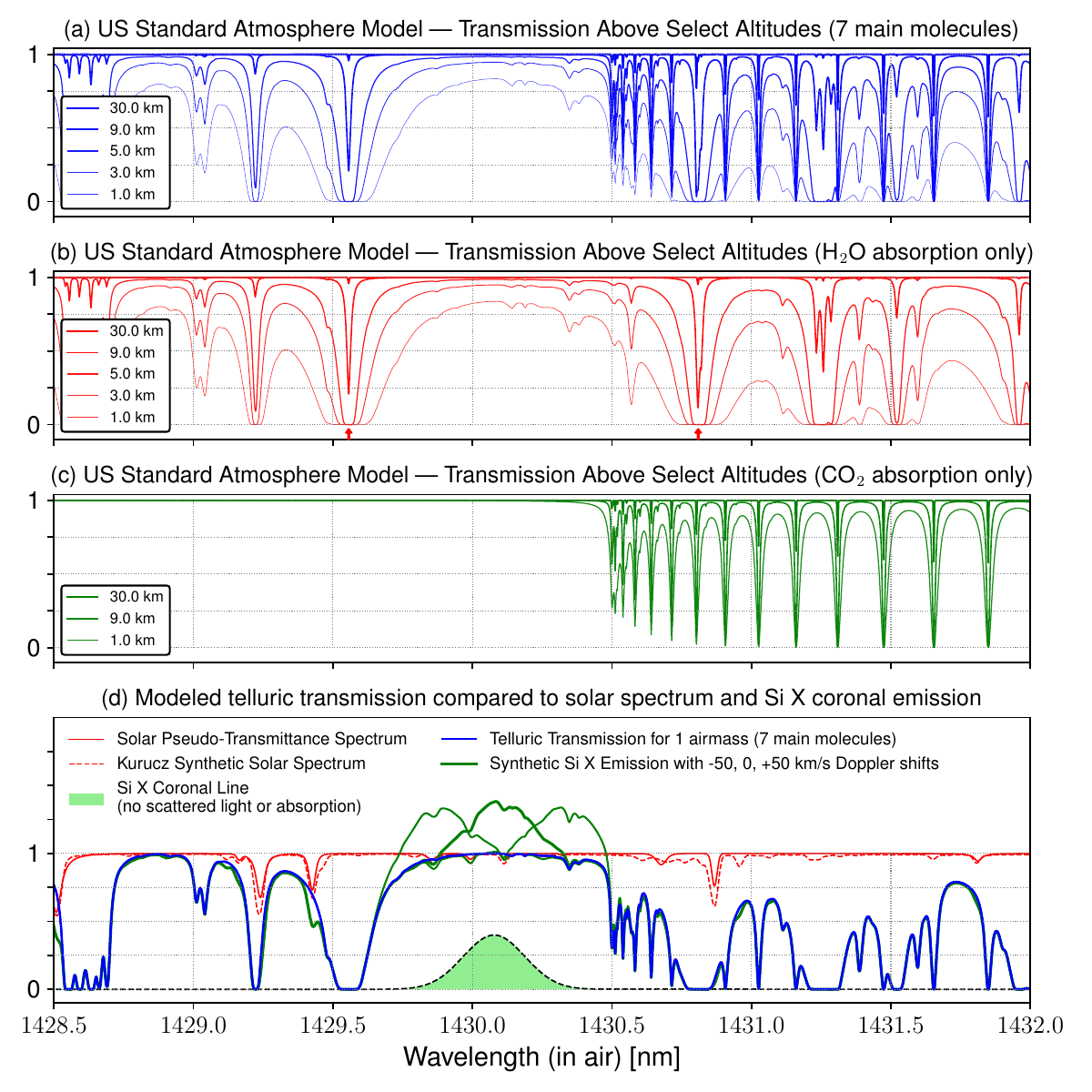} \\
\caption{
    \small
Synthetic telluric transmission spectrum near the \ion{Si}{10} 1430 nm coronal emission line.  (a) The fractional transmission spectrum for all seven most dominant molecular species in the Earth's atmosphere for the US standard reference model at select altitudes.  (b) The isolated H$_{2}$O component of the transmission spectrum.  (c) The CO$_{2}$ contribution to the transmission. (d) Synthetic coronagraphic spectra (green lines) and its constituent components as labeled and described in the text.
\vspace{-5mm}
}
\label{fig:telluricAbs}
\end{figure*}

The \ion{Si}{10} 1430 nm forbidden line (2s$^2$ 2p $^2$P$_{3/2 \rightarrow 1/2})$ is one of the brightest infrared lines emitted by the solar corona and has specific utility for inferring key properties of the coronal plasma.  Under coronal equilibrium conditions, it has a peak formation temperature near $10^{6.15}$ K. Its upper level can be excited both collisionally and radiatively from the ground state, resulting in density-sensitive emissivity that is also polarized through scattering in the Hanle-saturated regime \citep{judge1998,schad2020}.  On account of its long wavelength, it is a prime candidate for Zeeman-effect measurements of the ill-measured coronal magnetic field amplitude \citep{judge2001_techNote} and is consequently a key target for large-aperture coronagraphy at the Daniel K. Inouye Solar Telescope \citep[DKIST:][]{rimmele2020}. \citet{dima2019_spw8} and \citet{dima2020} propose using the \ion{Si}{10} line as part of multi-line inversions of coronal magnetic fields using full Stokes observations of multiple forbidden lines and/or joint linear-polarized observations of forbidden and permitted lines. Still, despite these advantages, \ion{Si}{10} 1430 nm observations are sparse, especially outside of solar eclipses, in part due to the challenges introduced by variable atmospheric transmission bands that dominate this spectral window (see Figure~\ref{fig:telluricAbs} and also \citet{ali2022}).

Eclipse observations acquired from high altitude aircraft help to minimize the impact of telluric absorption and have led to many discoveries of infrared coronal lines.  Using this approach, \citet{mangus1965} reported the first likely detection of the \ion{Si}{10} line, which was confirmed a year later by \citet{munch1967}.  Their reported observed wavelength ($1430.5 \pm 0.4$ nm in air), which was consistent also with follow-up observations by \citet{olsen1971}, located it on the edge of a narrow terrestrial transmission window, leading these authors to suggest that at least half of the line would be observable from the ground without severe telluric attenuation.  Later, \citet{penn1994_siX} successfully measured the line coronagraphically (\textit{i.e.} not during an eclipse) from an elevation of 2823 m and found a slightly shorter wavelength ($1430.078 \pm 0.008$ nm in air).  At this wavelength, the line lies at the center of a narrow (${\sim}0.85$ nm wide) transmission window, which in principle allows full detection of the ${\sim}0.25$ nm wide (full width at half maximum, FWHM) coronal line.  Theoretical atomic structure calculations by \citet{liang2012} give the transition's upper level energy to be 6990.5952 cm$^{-1}$, yielding a theoretical air-equivalent center wavelength of 1430.1024 nm.   Recent high-altitude eclipse observations by the Airborne Infrared Spectrometer \citep{samra2022} further corroborate the shorter center wavelength value, as do the linear-polarized coronagraphic observations by \citet{dima2019} and the first reported coronal spectra from DKIST \citep{schad2023}.

Here we present observations acquired by DKIST during instrument commissioning activities that provide the first spatially resolved measurements of \ion{Si}{10} 1430 nm across an active region.  Given the challenges posed by telluric absorption, we begin in Section~\ref{sec:telluric} with a review of the constituent terrestrial absorbers, which inform the employed calibration strategies.  The observations are then introduced in Section~\ref{sec:obs}, followed by a description of the data processing and analysis methods in Section~\ref{sec:dataProcessing}, and finally a discussion of our results in Section \ref{sec:results}.

\section{Telluric Absorption Near 1430 nm}\label{sec:telluric}

To study the role of telluric absorption near the \ion{Si}{10} 1430 nm line, we use the HITRAN molecular spectroscopic database, together with reference terrestrial atmospheric profiles, to synthesize the absorption of the seven dominant molecular species in the Earth's atmosphere:  H$_{2}$O, CO$_{2}$, O$_{3}$, N$_{2}$O, CO, CH$_{4}$, and O$_{2}$ \citep{gordon2022}.  For this exercise, we make use of the Py4CAtS (PYthon for Computational ATmospheric Spectroscopy) code \citep{schreier2019} as well as the US Standard atmosphere model of \citet{anderson86}, which provides temperature, pressure and constituent volume mixing ratios as a function of altitude.  The precipitable water vapor in this model is 14.4 mm; meanwhile, we scale the CO$_{2}$ ratios to a vertical column density of 420 parts per million by volume, in accordance with recent global averages.

Figure~\ref{fig:telluricAbs} panel (a) plots the telluric transmission as a function of wavelength for one airmass at select altitudes (heights above sea level) including all seven molecules.  Between 1428.5 and 1432 nm, the telluric absorption is dominated by H$_{2}$O and CO$_{2}$, whereas O$_{3}$, N$_{2}$O, and CH$_{4}$ transitions contribute less than 1\%. In panels (b) and (c), the separated components of cumulative H$_{2}$O and CO$_{2}$ absorption are displayed as a function of height\textemdash DKIST is located at 3067 m.  The large strength of many of the H$_{2}$O and CO$_{2}$ transitions lead to strong absorption even at high altitudes followed by continued wing broadening at lower heights.   The strong H$_{2}$O transitions at 1429.56 and 1430.81 nm (marked with arrows in panel b) strongly influence the shape of the narrow transmission window where the \ion{Si}{10} 1430 nm coronal line forms.  Both saturate at altitudes above 5km, implying that, at DKIST, the relationship between the total vertically-integrated water column and the absorbance is non-linear (\textit{i.e.}, it deviates from the Beer-Lambert law).  Meanwhile, additional strong CO$_{2}$ absorption occurs for wavelengths longer than 1430.5 nm, which also become saturated.

A coronagraphic observation of \ion{Si}{10} will include the coronal line and coronal continuum as well as some degree of atmospheric (telluric) and/or instrumental-scattered photospheric light.  Panel (d) of Figure~\ref{fig:telluricAbs} presents synthetic coronagraphic spectra (green lines) that include scattered light.  Here, we do not explicitly treat the sources of the telluric scattered solar light, which are likely dominated by aerosols near the Sun; nor, do we explicitly treat instrumental scattering sources dominated by dust.  We assume, as a first approximation, that the scattered photospheric light experiences the same telluric attenuation as the coronal signal.  The synthetic extraterrestrial solar spectrum from \citet{kurucz1994}, shown in red, is adopted for this demonstration, and the ratio of the coronal line peak amplitude to the scattered light peak amplitude is set to 0.4.  The \ion{Si}{10} rest wavelength is adopted from \citet{liang2012}, and we show two additional profiles with the coronal line Doppler shifted by $\pm50$ km s$^{-1}$.  Clearly, the presence of the scattered photospheric lines and telluric absorption influences the shape of the measured coronal line, and these effects are amplified when the ratio of the coronal line signal to the scattered light decreases.  

To extract reliable measurements of the coronal line properties in the presence of strong telluric absorption, careful treatment is required to separate the science signal from the background scattered light and telluric absorption. Not only does the telluric absorption shape the non-flat continuum near the coronal line, it also may attenuate a portion of the coronal line itself. In astronomy, multiple telluric calibration strategies have been advanced \citep[see, \textit{e.g.},][]{smette2015}, which also have similar applicability to the solar case.  These can include the use of atlases of telluric absorption spectra; calibration measurements of standard sources, which for the sun is the disk-center or integrated quiet sun spectra; or synthetic modeling of the telluric absorption. The targeted bandpass and stability of the atmosphere often dictate the method employed.  Below, we develop a method for DKIST observations of \ion{Si}{10} 1430 nm that can handle moderate variations in the total water column depth due to airmass or humidity changes without comprehensive radiative transfer modeling of the terrestrial atmosphere. 

%%%%%%%%%%%%%%%%%%%%%%%%%%%%%%%%%%%%%%%%%%%%%%%%%%%%%%%%%%%
\section{Observations} \label{sec:obs}

\subsection{DKIST/Cryo-NIRSP}

Off-limb observations of the \ion{Si}{10} 1430 nm coronal line were obtained by the \textit{Cryogenic Near-Infrared Spectropolarimeter} \citep[Cryo-NIRSP:][]{fehlmann2022} on 4 March 2022 during an instrument commissioning period.  Cryo-NIRSP's grating-based cryogenically-cooled spectrograph conducted a single raster scan of an active region located on the north east solar limb.  The telescope boresight (\textit{i.e.}, the target coordinates of the optical axis) was centered at $\left <X,Y\right >$ = $\left <-927'', 503'' \right >$ in helioprojective coordinates, which is 1.09 solar radii (R$_{\odot}$) from disk center.  The 5 arcminute limb occulter mechanism was deployed at the Gregorian focus to occult the near limb during the observation (see location in Figure~\ref{fig:sdo_dem}).   Cryo-NIRSP conducted a 201 position raster scan of the solar image across its $0.5''$ wide, $230''$ long slit, which was oriented radially relative to the Sun.  A raster step size of $1''$ resulted in a total field of view (including the occulted field) of ${\sim}200'' \times 230''$.  The spatial sampling along the slit is $0.12''$ pixel$^{-1}$, and the spectral dispersion is 4.88 pm pixel$^{-1}$ over the observed range from 1428.1 to 1432.4 nm in air-equivalent wavelengths.  The spatial resolution is seeing-limited (${\gtrsim}1''$) while the spectral resolving power R is approximately 45,000. 

\begin{figure}
\centering
\includegraphics[width=0.95\columnwidth,clip=]{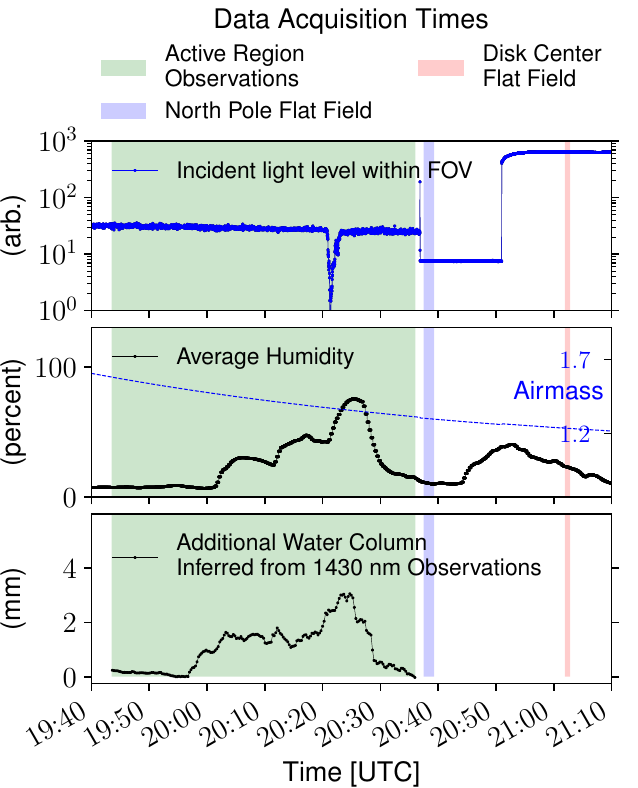} \\
\caption{(top) Auxiliary measurements of the incident light level (on a log scale) measured by the DKIST Target Acquisition Telescope at a wavelength of 656 nm over an 5 arcminute field-of-view approximately centered along the pointed DKIST optical axis.  Large changes correspond to pointing changes of DKIST (esp. off limb and on disk) and/or local intrusion by clouds. (middle) Auxiliary measurements of the average external humidity. (bottom) Inferred additional water column depth in 1430 nm observations compared to flat-field data (see Section~\ref{sec:background_removal}).
\vspace{-4.5mm}
}
\label{fig:humidity}
\end{figure}

At each step position, two repeats of an 8-state polarimetric modulation sequence were executed, resulting in 16 acquired exposure readout sequences (also known as `ramps' for the up-the-ramp sampling mode of the detector).  Each individual exposure sequence consisted of a frame reset followed by 10 non-destructive readouts of the H2RG detector with a 90.4 msec frame time.  The last 8 readouts, corresponding to a total of 723.2 msec of integration, are used to determine the incident flux.  The raster scan started at 19:43:29 UT, continuing for ${\sim}$52 minutes, until 20:36:04 UT, during which time the telescope elevation angle traversed from 39$^{\circ}$ to 50$^{\circ}$. For the purposes of flat-fielding and scattered-light correction, a set of exposures was obtained immediately after the science observations (as depicted in Figure~\ref{fig:humidity}) at a single slit position targeting the north solar pole, centered at 1.25 R$_{\odot}$, in a coronal hole where the line emission is comparatively negligible.  Disk center calibration observations were also obtained at 21:01 UT after first deploying an attenuation filter with an optical density of 4.5 (\textit{i.e.}, a transmission of 0.00316\%) at 1430 nm.  These data are used for photometric calibration relative to the disk center spectral radiance.

Qualitatively, the circumsolar scattered light on this date was sufficiently low for coronagraphic spectral measurements; however, the terrestrial atmosphere was not stable throughout the observations.  The top panel of Figure~\ref{fig:humidity} shows auxiliary measurements of the incident light level, which is measured at 656 nm by the Target Acquisition Telescope (TAT) and averaged over a 5 arcminute diameter centered on the DKIST boresight; while, the middle panel shows an aggregate value of the average external humidity.  Approximately 15 minutes (50 raster steps) into the science observations, the humidity markedly rises, yet the TAT light level is not significantly affected until between 20:20 and 20:23 UT when small clouds briefly traverse the telescope's line of sight.  The subsequent north pole flat field observations were acquired after the rastered science observations during a period of relatively lower humidity and lower airmass (calculated as the secant of the zenith angle).  The consequences of these unstable conditions are reflected in our calibration approach discussed below in Section~\ref{sec:dataProcessing}.

\subsection{SDO/AIA}

We analyze these DKIST/Cryo-NIRSP observations jointly with observations acquired by the Solar Dynamics Observatory's (SDO) \citep{pesnell2012} Atmospheric Imaging Assembly \citep{lemen2011}.  We make use of all level 1 EUV data products in the 94, 131, 171, 193, 211, and 335 \AA\ bandpasses with a pixel size of $0.6''$.  The nominal cadence of these observations is 12 seconds per channel.   For quantitative analysis and comparison with the DKIST observations, we average the data in each channel over the 2 hour time period centered on the DKIST/Cryo-NIRSP observation (19:40 to 20:40 UT) as described in Section~\ref{sec:coalign}.

%%%%%%%%%%%%%%%%%%%%%%%%%%%%%%%%%%%%%%%%%%%%%%%%%%%%%%%%%%%
\section{Methods} \label{sec:dataProcessing}

\subsection{Cryo-NIRSP Data Processing} 

The calibration steps applied to these Cryo-NIRSP data are similar to those presented by \citet{schad2023}.  The initial steps start with detector calibrations, including non-linear response correction, up-the-ramp flux fitting, and dark current subtraction.  Here, we do not analyze the data polarimetrically; instead, all modulation states are averaged during data reduction.  Also, we elect to use only one of the two orthogonally-polarized beams that exist in the data.  The flat-field response of the instrument is partially addressed in two steps.  First, a map of the pixel-to-pixel detector gain variations (\textit{i.e.,} as introduced by different responsitivities of the detector pixels) is created by removing the spectral features from the north pole limb observations.   This involves median filtering the data spatially along the slit with a boxcar kernel of 3 pixels and then applying a 3rd order Savitzky-Golay filter with a 7 pixel window length in the spectral direction.  The resulting smoothed spectra is removed from the gain image via division to create the pixel-by-pixel gain map that is applied to the data frames.  The second portion of the flat-field correction removes a linear slope in the spectral throughput of the instrument, which is inferred through least-squares fitting of the average observed spectrum to the NOAO/NSO FTS solar and telluric atlas spectrum \citep{livingston1991}.  Both the solar and telluric atlas spectra are modified during this process to infer the Cryo-NIRSP wavelength dispersion axis, the approximate resolving power, as well as the spectral transmission slope.  Despite the large opacity of many of the telluric transitions, we scale the telluric atlas absorbance during the fitting according to the Beer-Lambert law, which is adequate at this stage of the processing.  The grating induced spectral curvature (or smile) is calibrated and removed via linear interpolation.  Finally, the data is photometrically calibrated by dividing the observed spectral radiance by that measured in the disk center continuum along the slit during the disk center calibration observations.  In this step, we account for both the attenuation filter's optical density and differences in exposure time.  The relative spectral radiance in units of millionths of the disk center radiance ($\mu B_{\odot}$) is used as a measure of the relative coronal brightness.  Meanwhile, conversion to absolute radiance units is done by multiplying by the cataloged value of the solar spectral continuum radiance at 1430 nm, \textit{i.e.} 9.06 $\times 10^{7}$ photons cm$^{-2}$ arcsec$^{-2}$ nm$^{-1}$ from Allen's Astrophysical Quantities \citep{cox2000}.

\begin{figure*}
\centering
\includegraphics[width=0.95\textwidth,clip=]{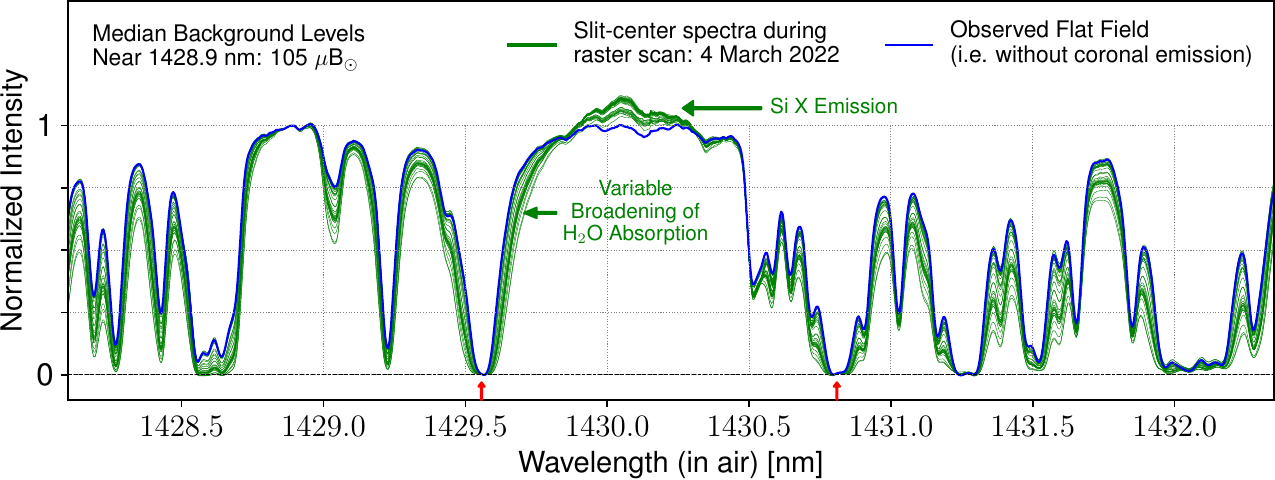} \\
\caption{Comparison of observed coronal spectral profiles obtained from the middle of the slit for all 201 raster step positions (green lines) with the "flat"-field spectrum acquired at the North Pole (blue line).  Each observed profile has been normalized to its value near 1428.95 nm.  This normalization re-scales the data, which originally has a median background flux of $105~\mu B_{\odot}$. Excess emission at 1430 nm corresponds to the coronal emission line.
\vspace{-4.5mm}
}
\label{fig:spectraDiversity}
\end{figure*}

\subsection{Background removal and line fitting}\label{sec:background_removal}

Returning to the issue of background removal, \textit{i.e.} the separation of scattered light and telluric absorption from the observed coronal spectrum, we first illustrate in Figure~\ref{fig:spectraDiversity} the range of telluric absorption experienced during the 201 step raster scan.  For each raster step, we normalize the observed spectra at the middle of the slit by the continuum scattered light amplitude near 1428.95 nm (green lines). This normalization facilitates comparison with the observed flat field spectrum (blue), which corresponds to the off-limb solar north pole observations with the above partial gain corrections. The median scattered light amplitude for this position is 105 $\mu B_{\odot}$.\footnote{These data were acquired during instrument and facility commissioning.  Scattered light levels are higher compared to expectations for steady-state operations by at least a factor of 2.} The coronal line appears as excess emission near 1430 nm, strongly blended with the telluric and background solar photospheric lines, with an amplitude of approximately 10\% of the scattered light magnitude (\textit{i.e.}, ${\sim}10~\mu$B$_{\odot}$).  Variability in the telluric absorption near the coronal line is most pronounced in the wings of the strong, saturated, H$_{2}$O absorption line centered at 1429.56 nm.  In contrast, variations in the CO$_{2}$ absorbance between the data and flat-field are as not readily apparent over the 0.5 change of airmass during the data acquisition, especially as these lines are heavily blended with the water lines.

\subsubsection{Modeling the observational components}

To facilitate background corrections, we develop an empirical model of the multiple components contributing to the observed spectrum.  Our objective is to extract the coronal emission line properties while avoiding detailed radiative transfer modeling of the atmospheric scattered light and the atmospheric extinction of all molecular species.  For $\hat{I}_{obs}(\lambda)$, defined as the observed coronal spectrum normalized by the solar continuum intensity at disk center ($I_{c}$), we may consider a model where
\begin{align}
    \hat{I}_{obs}(\lambda) & \simeq \frac{G(\lambda)}{I_{c}} \left [ I_{E}(\lambda) + I_{K} + I_{scattered}(\lambda) \right ] e^{-\tau_{\earth}(\lambda)}  
\end{align}
where $I_{E}$ is the coronal emission line radiance, $I_{K}$ is the Thomson-scattered K-corona radiance (absent of spectral structure), $I_{scattered}$ is the atmospheric/instrumental scattered light profile, and $\tau_{\earth}(\lambda)$ is the wavelength-dependent optical thickness of the telluric absorption at the time of the observation. The additional term $G(\lambda)$ represents a spectral-dependent optical response function (\textit{i.e.} gain) that accounts for residual flat-fielding artifacts in the reduced data.  In this case, this includes interference fringing with an ${\sim}$0.34 nm period (${\sim}70$ spectral pixels) and $\pm1\%$ amplitude that show relative phase shifts along the slit. These fringes are not otherwise removed during data processing and are difficult to remove with other methods \citep[see, \textit{e.g.},][]{schad2023} due to the crowded nature of this spectral window. 

The scattered light profile $I_{scattered}$ is photospheric in origin; however, it is introduced by the extended wings of the combined point-spread-function of the terrestrial atmosphere, the telescope, and the instrument, which is complex to model and measure.  Given the weak center-to-limb variation of the photospheric lines near 1430 nm,  we can approximate the magnitude of $I_{scattered}$ as some multiple $b$ of the disk center solar spectral radiance $I_{\odot}$, \textit{i.e.},
\begin{align}
    \hat{I}_{obs}(\lambda) & \simeq \frac{G(\lambda)}{I_{c}} \left [ I_{E}(\lambda) + I_{K} + bI_{\odot}(\lambda) \right ] e^{-\tau_{\earth}(\lambda)}.\label{eqn:iobs_approx}    
\end{align}
To help remove the influence of the gain term $G(\lambda)$, we note that the flat-field spectra normalized by $I_{c}$ can be approximated as 
\begin{align}
    \hat{I}_{f}(\lambda) & \simeq \frac{G(\lambda)}{I_{c}} I_{\odot}(\lambda) e^{-\tau_{f}(\lambda)} 
\end{align}
where $\tau_{f}(\lambda)$ is the telluric optical thickness at the time of the flat field.  Here, we assume the gain term is stable during the science and flat-field observations, which is not guaranteed; though, it is a fair approximation in this case.  Meanwhile, the relationship between the telluric optical thickness during the observation and during the calibrations can be written as $\tau_{\earth} = \tau_{f} + \Delta\tau$; therefore,
\begin{align}
    \hat{I}_{f}(\lambda) & \simeq \frac{G(\lambda)}{I_{c}} I_{\odot}(\lambda) e^{-\tau_{\earth}}e^{\Delta\tau}.\label{eqn:iflat_approx}
\end{align}
By dividing Eq.~\ref{eqn:iobs_approx} by Eq.~\ref{eqn:iflat_approx}, we eliminate the gain term and obtain 
\begin{align}
    \hat{I}_{obs}(\lambda)
    & \simeq 
    \left [ 
    \frac{I_{E}(\lambda) + I_{K} + bI_{\odot}(\lambda)}{I_{\odot}(\lambda)e^{\Delta\tau}}
    \right ]
    \hat{I}_{f}(\lambda)
    \equiv
     \hat{I}_{model}(\lambda),
\end{align}
where $\hat{I}_{model}(\lambda)$ is defined as the pseudo-modeled equivalent of the observed off-limb spectrum. Rearranging the terms, and dividing both sides by $I_{c}$, we can also write
\begin{align}
\frac{I_{E}(\lambda) + I_{K}}{I_{c}} \simeq 
\left [ 
\frac{ \hat{I}_{obs} (\lambda) }{ \hat{I}_{f}(\lambda)e^{-\Delta\tau}}  - b 
\right ] 
\frac{I_{\odot}(\lambda)}{I_{c}}.\label{eqn:final_model}
\end{align}
The left side of this equation represents the coronal signals normalized by the disk-center intensity.  On the right side, the coronal observations $\hat{I}_{obs}(\lambda)$ are first divided by the flat-field adjusted for additional opacity at the time of the observation ($\hat{I}_{f}(\lambda)e^{-\Delta\tau}$), which removes the effects of the gain and telluric absorption from the observation. The scattered light fraction $b$ is then subtracted, and the effects of solar photospheric spectrum removed via multiplication. 

\subsubsection{The Objective Function for Least Squares Fitting}\label{sec:objective}

We use Equation~\ref{eqn:final_model} as our model to extract the \ion{Si}{10} line signals from the Cryo-NIRSP observations via least-squares fitting.  Our objective function is given by
\begin{align}
    \min \sum_{\lambda} &\left ( 
\left [ \frac{ \hat{I}_{obs} (\lambda) }{ \hat{I}_{f}(\lambda)e^{-\Delta\tau}}  - b \right ] 
\frac{I_{\odot}(\lambda)}{I_{c}} - \right.  \nonumber \\ 
& \left. \frac{I_{E}(\lambda) + I_{K}}{I_{c}} \right )^{2} \frac{\omega_{\lambda}}{\sum_{\lambda} \omega_{\lambda}} \label{eqn:obj1}
\end{align}
where $\omega_{\lambda}$ are spectral-dependent fitting weights.  Alternatively, the objective function can be written as
\begin{align}
    \min \sum_{\lambda}
\left [ \hat{I}_{obs} (\lambda) - \hat{I}_{model}(\lambda)  \right ]^{2} \frac{\omega_{\lambda}}{\sum_{\lambda} \omega_{\lambda}},\label{eqn:obj2}
\end{align}
which reduces the amplification of errors otherwise caused by the division of the highly structured flat field spectra in the Equation~\ref{eqn:obj1}.  That said, we do not explicitly treat the propagation of measurement uncertainties here. Instead, for the same weighting function $\omega_{\lambda}$, we can find consistent minima using either objective function.

To optimize the model given in Equation~\ref{eqn:obj1}, we assume the coronal line signal $I_{E}(\lambda)$ is well represented by a Gaussian profile. Meanwhile, the $I_{\odot}/I_{c}$ term is the normalized solar spectrum without telluric absorption.  For this we adopt the Solar Pseudo-Transmittance Spectrum (SPTS) derived from high spectral resolution disk center observations by the TCCON network\footnote{\url{https://mark4sun.jpl.nasa.gov/toon/solar/solar_spectrum.html}}, which provides a cleaner empirical representation of the solar features in this window than other available atlases (see Figure~\ref{fig:telluricAbs}).  The changes in telluric opacity are assumed to be dominated by additional water absorbance, \textit{i.e.} $\Delta\tau \approx \Delta\tau_{H_{2}O}(\lambda)$.  We further approximate that the additional water column absorbance can be modeled, using the HITRAN database, by a simple constant-property cloud (or slab) model whose parameters include the local pressure, temperature, water volume mixing ratio, and slab-thickness.  For numerical efficiency, we only include water lines with strengths larger than 10$^{-26}$ cm$^{-1}$ / (molecule m$^{-2}$) at 296 K, which includes 87 transitions between 1428.5 and 1430.4 nm over which we apply the model.  For the opacity calculations, we use Voigt line profiles and account for air- and self-broadening effects, the temperature dependent line strength, and pressure-induced line shifts, as per the line parameter definitions of the HITRAN database\footnote{\url{https://hitran.org/docs/definitions-and-units/}}. The total internal partition sums for molecules are sourced from \citet{gamache2021}.

In all, there are 12 free parameters in our model, including 3 Gaussian coronal line parameters, the K-coronal intensity $I_{k}$, the scattered light magnitude $b$, and the 4 parameters for the telluric cloud-model.  The last three parameters we introduce are a spectral shift for the additional telluric absorbance, a spectral shift for the SPTS solar spectrum, and a Gaussian width of the spectral line-spread-function.  Our goal is to reliably extract the 3 coronal line parameters.  We do not necessarily extract the other parameters reliably, especially as there are multiple degeneracies and correlations between them.  In Appendix~\ref{sec:appendix1}, we have investigated whether the model degeneracies affect the coronal line parameter estimation and have concluded that the effects are benign in this case.

\subsubsection{Fitting weights and optimization}

\begin{figure*}
\centering
\includegraphics[width=0.495\textwidth,clip]{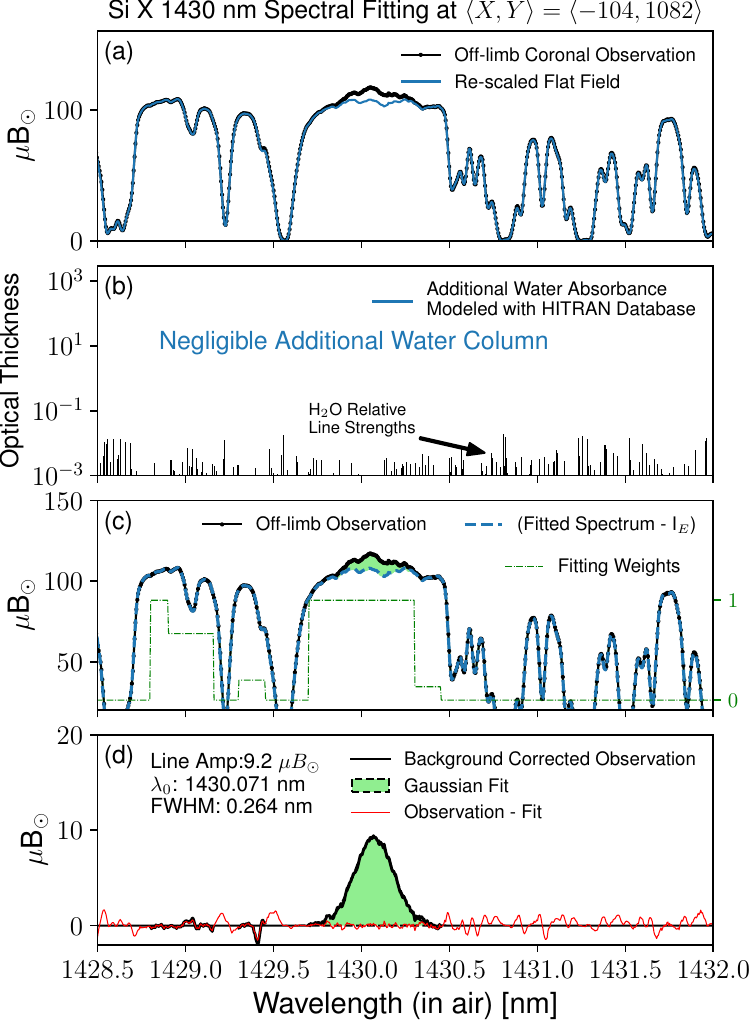}
\includegraphics[width=0.495\textwidth,clip=0 0 0 4cm]{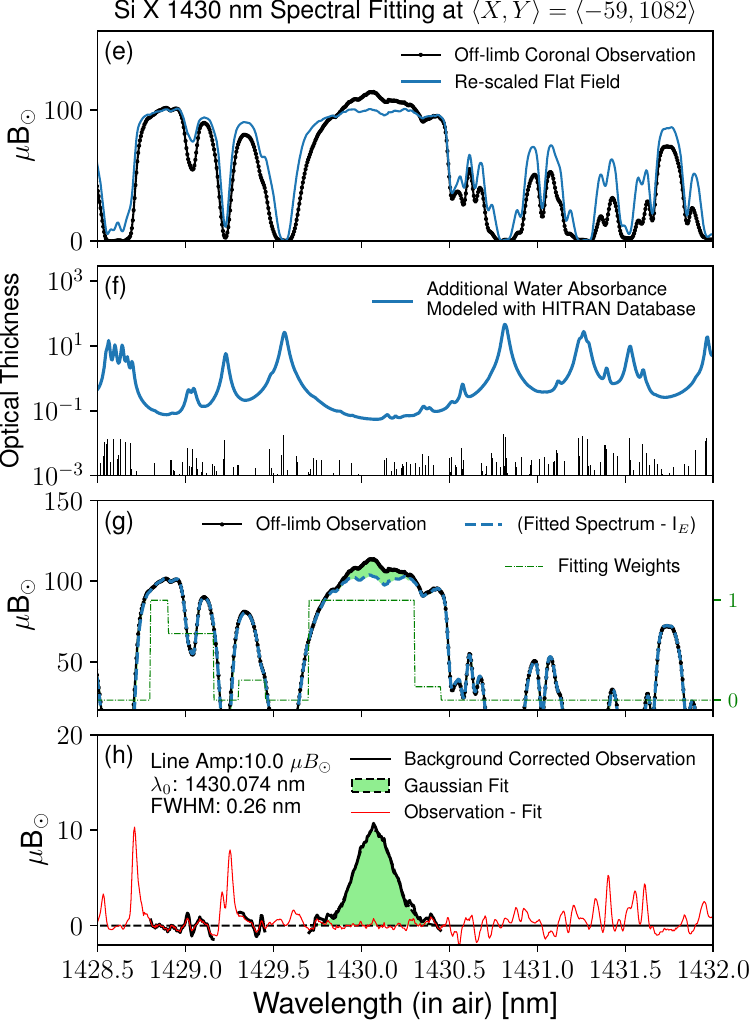}
\caption{Example model fitting of coronal spectra at 1430 nm to extract the coronal line properties for (a-d) the case of little additional water content compared to the flat-field and (e-h) the case with substantially more additional telluric water opacity.
\vspace{-4mm}
}
\label{fig:fitting_examples}
\end{figure*}

For the fitting weights $\omega_{\lambda}$, we define a piece-wise constant function that provides more weight to the fit near portions of the spectrum less affected by observational uncertainties (see green dashed lines in panels c and g of Figure~\ref{fig:fitting_examples}).  The highest weight is applied to the region around the coronal line (1429.7 to 1430.3 nm) as well as the high transmission window between 1428.8 and 1428.9 nm. The next highest weighting is applied between 1428.9 and 1429.16 nm where non-saturated telluric water absorption occurs in absence of strong photospheric line blends.  Lower weights are applied to the ranges of 1429.3 to 1429.45 nm and 1430.3 and 1430.45 nm due to being portions of the spectrum with blended telluric and solar contributions. Finally, the deep saturated cores of the telluric water lines, as well as the CO$_{2}$ absorption band ($\lambda>1430.45$ nm), are given a zero weight, and therefore do not contribute to the fit. 

To minimize the objective function (Equation~\ref{eqn:obj1}), the Differential Evolution method \citep{storn1997} for bounded global optimization is applied, along with a subsequent polishing of this initial solution using the Levenberg-Marquardt algorithm as implemented by \citet{more1978}.  Two examples of this model fitting, for observed spectra with substantially different additional water opacity $\Delta\tau_{H_{2}O}(\lambda)$ are shown in the left and right columns of Figure~\ref{fig:fitting_examples}. In the top panels, (a) and (e), the off-limb coronal observations are shown in comparison to the flat field spectrum linearly scaled to approximately match the continuum level.  Clearly, the coronal spectrum in panel (e) has a larger degree of telluric absorption compared to that of the flat field.  In panels (b) and (f), we plot the additional telluric optical thickness, $\Delta\tau \approx \Delta\tau_{H_{2}O}(\lambda)$, as inferred by our model. Also shown are the positions and relative line strengths (also on a log scaled vertical axis) of the water transitions.  In panel (b) the additional absorption is negligible.  Panels (c) and (g) compare the observations to the fitted spectrum with the inferred coronal line emission subtracted, \textit{i.e.} $\hat{I}_{model}(\lambda) -  I_{E}(\lambda)$.  This helps illustrate the portion of the observed signal attributed to the coronal emission (as highlighted in green).  Finally, in panels (d) and (h), we plot the background-corrected observations of the coronal emission line (thick black line) which, following from Equation~\ref{eqn:final_model}, are given by 
\begin{align}
    \left [ \frac{ \hat{I}_{obs} (\lambda) }{ \hat{I}_{f}(\lambda)e^{-\Delta\tau}}  - b 
\right ] 
\frac{I_{\odot}}{I_{c}} - \frac{I_{K}}{I_{c}}.
\end{align}
To remove the amplified errors caused by the small number division in $\hat{I}_{f}(\lambda)$), we only plot the values where $\omega_{\lambda}>0$. The Gaussian fit is shown overplotted, as well as the the residuals of $\hat{I}_{model}(\lambda) -  I_{E}(\lambda)$ (red lines).  Despite the relatively large swings of telluric absorption and the high degree of scattered light compared to the coronal line signal, we are able to adequately extract the coronal line signals with this method.  The presence of residual solar and telluric line features in the extracted coronal spectra is minimal. Larger fit residuals occur near the sharp edges and deep minima of the strongest telluric water lines; though, these do not substantially affect the coronal line in this case.

\begin{figure*}
\centering
\includegraphics[width=0.9\textwidth,clip]{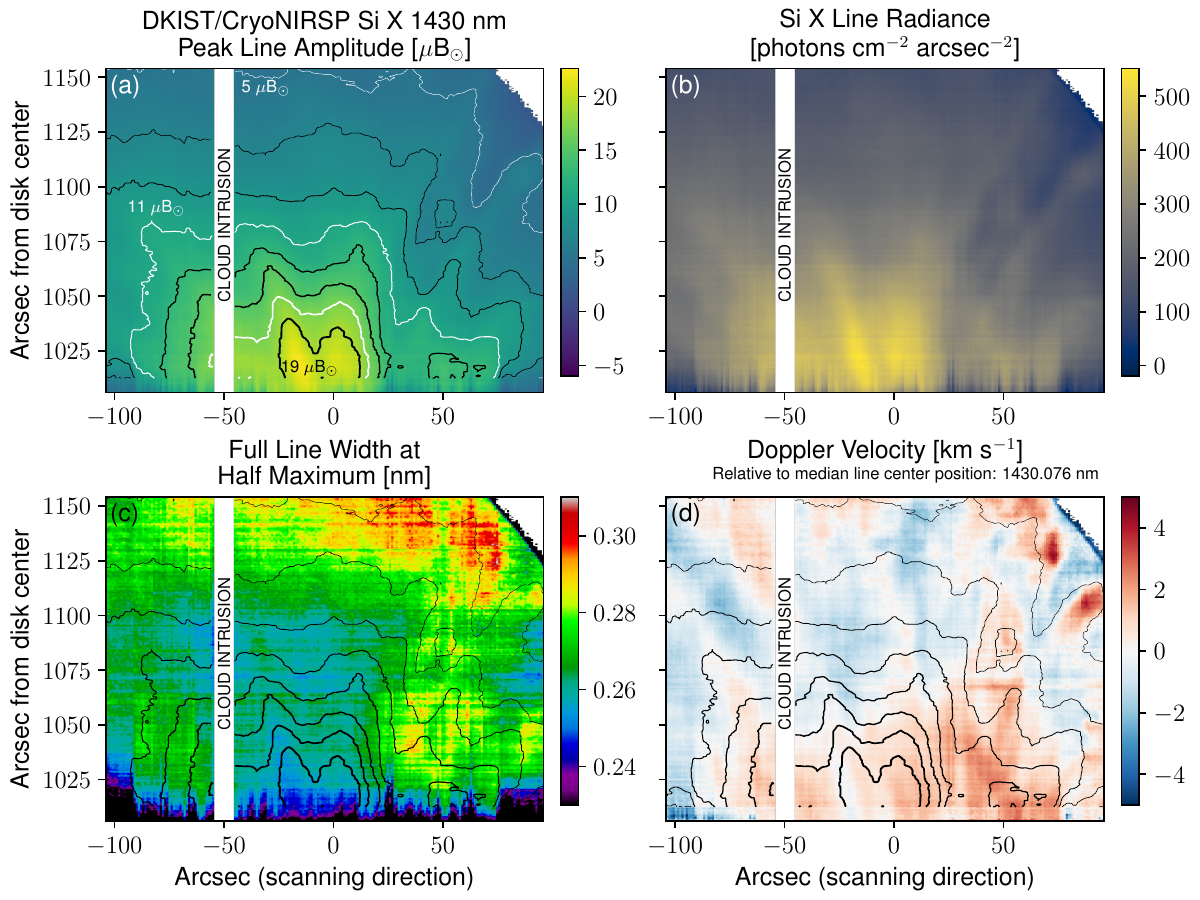}\\
\caption{Maps of the derived Gaussian line parameters for the \ion{Si}{10} 1430 nm observations on 4 March 2022 by DKIST/Cryo-NIRSP.  (a) The peak line amplitude given in units of millionths of the disk center intensity. (b) Total line radiance in photometric units. (c) The full width at half maximum of the line profile in nanometers.  (d) The Doppler Velocity inferred from the line center position relative to the median line position over the raster scan.   Contour lines of the peak line amplitude for equal 2$\mu$B$_{\odot}$ samples between 5 and 19 $\mu$B$_{\odot}$ are shown in panels (a), (c), and (d).
\vspace{-4mm}
} 
\label{fig:siX_maps}
\end{figure*}

The extracted coronal line parameters across the entire observed field-of-view are presented in Figure~\ref{fig:siX_maps}, which are discussed further below.  Interesting, in the lower panel of Figure~\ref{fig:humidity}, we also plot the median inferred additional water column depth (\textit{i.e.} the additional precipitable water vapor) as a function of time in the coronal raster scan, which expectedly correlates well with the average humidity measured external to the telescope.  At the moment of largest average external humidity, an additional ${\sim}3$ mm of vertically-integrated water column depth is inferred by our model. 

\subsection{Cryo-NIRSP coalignment with AIA}\label{sec:coalign}

To further analyze the extracted \ion{Si}{10} measurements, we manually co-align the Cryo-NIRSP observations with SDO/AIA EUV observations.  The accuracy of the spatial world coordinate information in these DKIST commissioning observations is limited.  To co-align, we compare the spatial structure in maps of the fitted \ion{Si}{10} line amplitude with that in a two-hour average of SDO/AIA 193 \AA\ data, for which the characteristic formation temperature---dominated by $10^{6.2}$ \ion{Fe}{12} emission \citep{odwyer2010}---most closely matches that of \ion{Si}{10}.  The plate scale and raster step size of Cryo-NIRSP is presumed correct as well as the angular orientation of the field relative to the slit.  Scalar shifts in the horizontal and vertical axes of the Cryo-NIRSP coordinates are applied until a satisfactory match of structures is achieved.  This process takes into account additional constraints such as the solar limb position and the approximately known offset of the occulter edge relative to the solar limb, as set by the telescope boresight coordinates.  Using this procedure, we estimate the coalignment of these data is good to within a few arcseconds, which is adequate for our purposes and will be improved upon in the future observations.  The coordinates given in Figures~\ref{fig:siX_maps} and~\ref{fig:sdo_dem} are the result of this coalignment process.

\subsection{SDO/AIA Differential Emission Measure (DEM) Analysis} 

\begin{figure}
\centering
\includegraphics[width=0.90\columnwidth,trim={0 0 0 -5mm},clip]{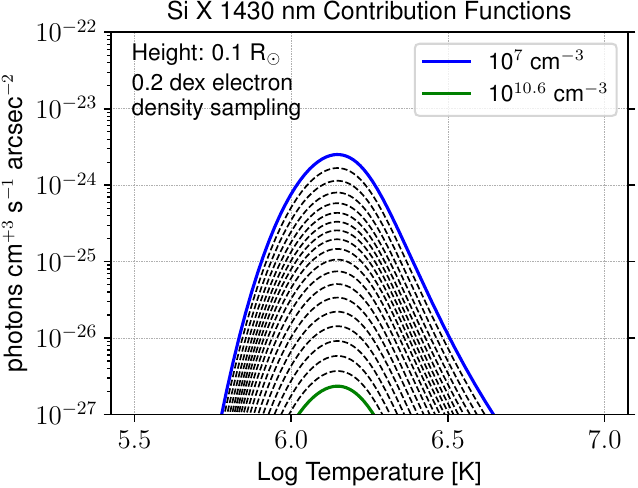}
\caption{Temperature-dependent contribution functions for \ion{Si}{10} 1430 nm at constant densities, sampled at 0.4 dex between $10^{7}$ and $10^{10.6}$ cm$^{-3}$.
\vspace{-5mm}
}
\label{fig:siX_contFnc}
\end{figure}

We apply a differential emission measure (DEM) analysis to the SDO/AIA EUV observations to extract the emission measure distribution as a function of temperature.  Assuming a unique relationship between density and temperature, the line-integrated intensity of a spectral line in the optically thin case is given by \citep{delzanna_mason2018} 
\begin{equation}
I_{ji}=\int_T G_{ji}\left(N_{\mathrm{e}}, T, ...\right) DEM(T){\ }dT \label{eqn:dem_G}
\end{equation}
where the contribution function $G_{ji}$ for a transition $j{\rightarrow}i$ is a function of temperature and density as well as elemental abundances and other applicable factors such as photoexcitation \citep[see, \textit{e.g.}, Eq. 5 of][]{schad2020}. $DEM(T)$ is the column differential emission measure that quantifies the amount of emitting plasma at a given temperature along an observed line-of-sight, \textit{i.e.} $DEM(T) = N_e N_H (\mathrm{d}h/\mathrm{d}T$). Similarly, the detected counts measured in a specific bandpass $i$ of AIA can be written as: 
\begin{equation}
y_{i} = \int_T R_{i}(T){\ }DEM(T){\ }dT 
\end{equation}
wherein the response function $R_{i}$ is a function only of temperature and abundances.  This relies on the contribution functions of the spectral lines forming in the AIA bandpasses being weakly dependent on the coronal density, which is not the case for the \ion{Si}{10} 1430 nm forbidden line whose contribution function are strongly density dependent (see Figure~\ref{fig:siX_contFnc}).  Using the set of six AIA EUV measurements, $DEM(T)$ can be inferred using various inversion strategies.  We use the fast simple algorithm presented in \citet{plowman2020} to reconstruct $DEM(T)$ using the 2 hour averages of AIA EUV data.  The recommended default input parameters are used together with the temperature response functions of the AIA bandpasses for the observation date and using coronal abundances from \citet{feldman1992}.  The response function calculations use version 9.0.1 of the {\sc Chianti} atomic database \citep{dere2019}. We find the inversion of the DEM converges well, and the mean relative errors in the reconstructed EUV 94, 131, 171, 193, 211, and 335\AA\ intensities are 30, 28, 7, 3, 3, and 24\%, respectively, across the FOV of Cryo-NIRSP.  A map of the relative error fraction for the reconstructed 193\AA\ emission is shown in Figure~\ref{fig:sdo_dem} panel (d).   Meanwhile, the inverted map of the DEM at $10^{6.15}$ K, near the formation temperature of \ion{Si}{10} 1430 nm, is shown in panel (c) of Figure~\ref{fig:sdo_dem}.  

\subsection[Density estimation using Si X radiances and the DEMs]{Density estimation using \ion{Si}{10} radiances and the DEMs}\label{sec:densityMethod}

\begin{figure*}
\centering
\includegraphics[width=0.95\textwidth]{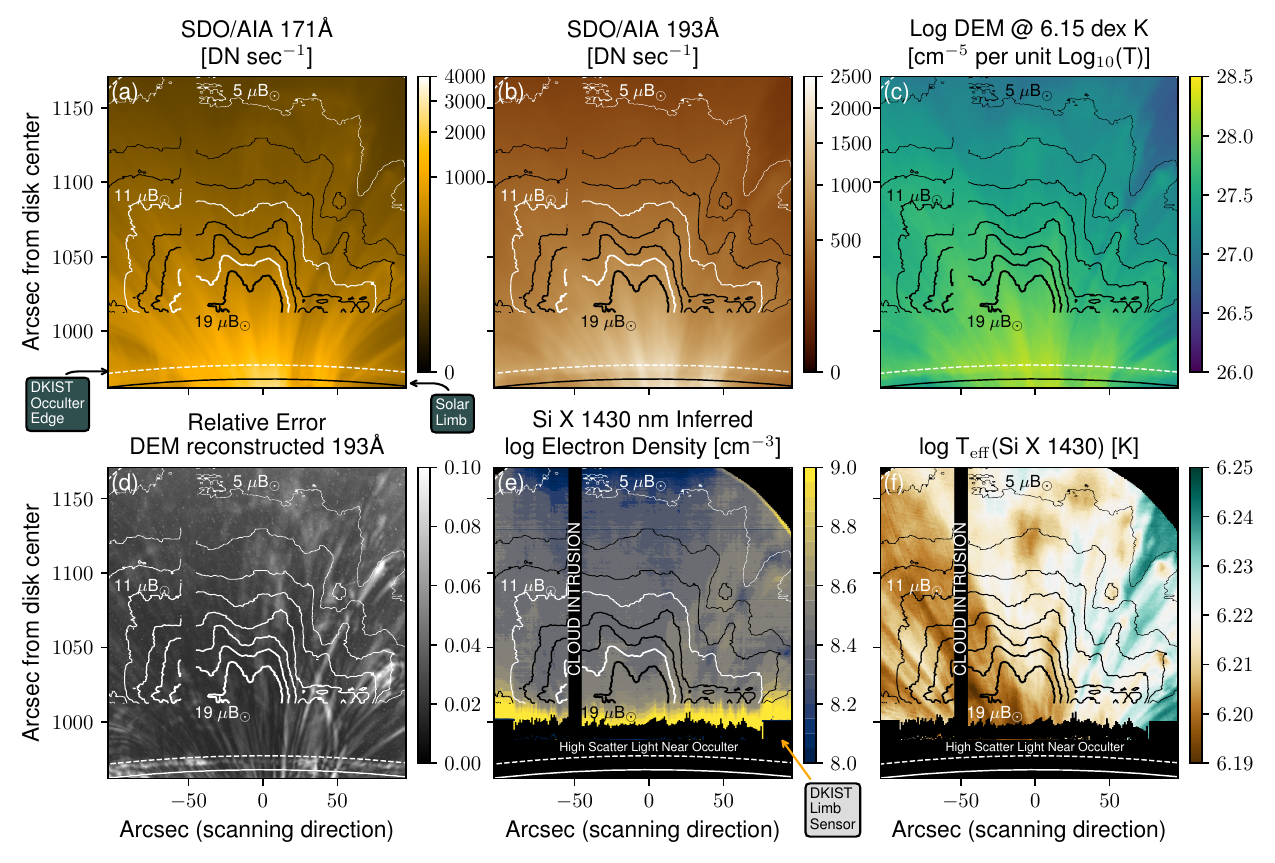}
\caption{Two hour averages of (a) SDO/AIA 171 \AA\ and (b) SDO/AIA 193 \AA\ data coaligned with the DKIST/Cryo-NIRSP \ion{Si}{10} 1430 nm observations.  (c) Inverted value of the DEM at $10^{6.15}$ K near the peak formation temperature of \ion{Si}{10}.  (d) Relative error of the intensity of the 193\AA\ channel as reconstructed by the derived $DEM$ (e) Inferred electron density for \ion{Si}{10} using the observed line radiance and the reconstructed $DEM$. (f) Effective temperature for \ion{Si}{10} calculated using the $DEM$ and the inferred electron density. Contour lines of the peak \ion{Si}{10} line amplitude for equal 2$\mu$B$_{\odot}$ samples between 5 and 19 $\mu$B$_{\odot}$ are shown in all panels, which are the same contour lines as in Figure~\ref{fig:siX_maps}. The solar limb and DKIST occulter positions are shown with solid and dashed line lines.
\vspace{-4mm}
}
\label{fig:sdo_dem}
\end{figure*}

In principle, with knowledge of the $DEM$ from the above reconstruction, one can forward synthesize the intensity of another spectral line using Equation~\ref{eqn:dem_G} and the line's respective contribution function, provided other variables entering the contribution function are known or can be constrained. Alternatively, with additional observations, one may infer other variables of the coronal plasma.  Here, we derive an estimate for the density along the line-of-sight that contributes to the \ion{Si}{10} 1430 nm line.  We first assume the Silicon abundance is constant with a standard normalized value of 8.1 dex as per the coronal abundances given in \citet{feldman1992}.  The corresponding temperature and density dependent contribution functions are calculated using the pyCELP package \citep{schad2021_pyCELP}, as shown in Figure~\ref{fig:siX_contFnc}.  These calculations include photoexcitation at a height of 0.1 solar radii above the solar surface.  The influence of atomic alignment on the intensity contribution function, as discussed Section 2.2.3 of \citet{schad2020}, is ignored here; for active region densities, atomic alignment could modify the contribution function by 5 to 10\% depending on the magnetic field orientation.  Since the contribution functions are strongly peaked in temperature, we consider a case of constant electron density along the line-of-sight.  This assumption, along with the many assumptions inherent to the DEM inversion technique and the coarse spatio-temporal alignment between the DKIST and SDO observations here, are fundamental limitations that we recognize here.  Accepting these limitations for this demonstration, we can derive an estimate for the electron density by determining the density value at which the line contribution function and the reconstructed $DEM(T)$ from Equation~\ref{eqn:dem_G} best matches the intensity observed by Cryo-NIRSP.  The corresponding simple objective function can be written as
\begin{equation}
\min \left |  \int_T G_{\tiny{Si X}}\left(N_{\mathrm{e}}, T \right){\ }DEM(T){\ }dT - \int_{\lambda} I_{E}(\lambda){\ }d\lambda  \right |
\end{equation}
for which only $N_{e}$ is a free parameter.  The result of this process for the DKIST/Cryo-NIRSP observations is shown in Figure~\ref{fig:sdo_dem} panel (e). 

\subsection[Effective Si X Temperatures]{Effective \ion{Si}{10} Temperatures}

From the DEM analysis, one may also derive an effective temperature for a particular line or bandpass via Eq. 93 of \citet{delzanna_mason2018}, \textit{i.e.}
\begin{align}
    \log T_{\text{eff}}=
    \frac{\int G_{\lambda}\left(N_{\mathrm{e}}, T, ...\right)~DEM(T)~\log T~dT}
    {\int G_{\lambda}\left(N_{\mathrm{e}}, T, ...\right)~DEM(T)~dT},
\end{align}
which calculates the $DEM$ weighted average of the temperature along the line of sight.  This approach offers an improved estimate for the line temperature compared to the assumption that a spectral line forms at its peak ionization fraction temperature in the coronal approximation, which has been adopted often in the past. Typically, this equation is applied in the case of EUV lines, for which the density dependence of the contribution is often negligible. This implies that the effective temperature is determined primarily by the $DEM$ itself and not variations of density (or abundance) along the line of sight.  We can calculate an effective temperature for the \ion{Si}{10} 1430 nm line by again adopting the perhaps crude assumption that the density is constant along the line-of-sight, which is derived using the methods in the previous section (\S\ref{sec:densityMethod}).  Given the derived density, the $DEM$, and the contribution functions, we determine $T_{\text{eff}}$ for \ion{Si}{10}, as shown in panel (f) of Figure~\ref{fig:sdo_dem}. 

\subsection[Si X 1430 nm Non-Thermal Velocities]{\ion{Si}{10} 1430 nm Non-Thermal Velocities}\label{sec:nonthermal}

\begin{figure}
\centering
\includegraphics[width=0.9\columnwidth,trim={0 0 0 -5mm},clip]{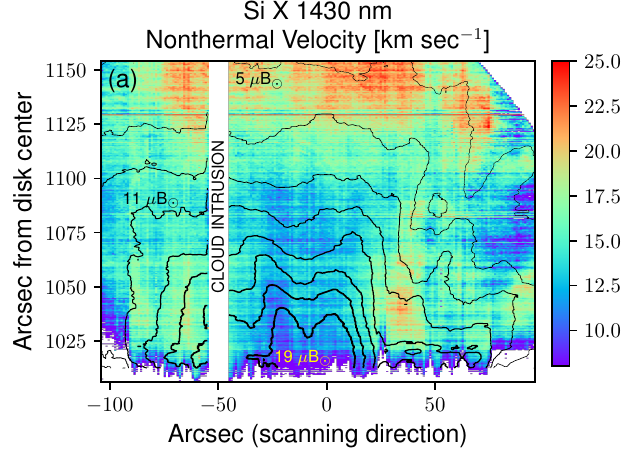} \\
\includegraphics[width=0.9\columnwidth,trim={0 0 0 -5mm},clip]{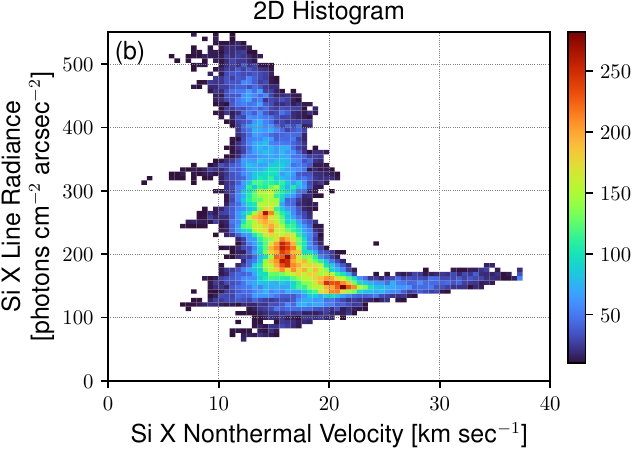} 
\caption{(a) Inferred nonthermal line width in velocity units for the \ion{Si}{10} 1430 nm line based on the effective temperature inferred from the SDO/AIA DEM analysis and the Cryo-NIRSP observations. (b) 2D histogram of \ion{Si}{10} line radiance versus nonthermal velocity.
\vspace{-7.5mm}
}
\label{fig:nonthermal}
\end{figure}

Finally, using the estimated effective temperature for \ion{Si}{10} across the observed field-of-view and the measured \ion{Si}{10} line widths, we may also calculate the nonthermal velocity $\xi$ component of the observed line width defined as
\begin{align}
\xi = 
\sqrt{
\frac{FWHM^2 - w_{\mathrm{I}}^2}{4 \ln 2\left(\frac{\lambda_0}{c}\right)^2}  - \left(\frac{2kT}{M} \right)}
\end{align}
where $w_{\mathrm{I}}$ is the instrumental spectral point-spread-function FWHM ($\approx\lambda/$R where the resolving power R${\sim}45000$), k is Boltzmann's constant, $T$ is the ion temperature (assumed here to be ${\approx}T_{\text{eff}}$), and M is the ion mass. Figure~\ref{fig:nonthermal} shows the map of the nonthermal velocity across the Cryo-NIRSP field-of-view as well as the 2D histogram of the total line radiance versus nonthermal velocity.  We note that the nonthermal line widths would scale upwards if we use the peak ionization fraction temperature of 10$^{6.15}$ K instead of the effective temperature $T_{eff}$ shown in panel (f) of Figure~\ref{fig:sdo_dem}, which has median value of the effective temperature 10$^{6.21}$.  That said, we note this is only 0.06 dex higher than the peak ionization fraction temperature, which is ${\sim}30\%$ of the FWHM of the contribution function versus temperature. The difference in the estimated nonthermal velocities may therefore be inconsequential relative to the inherent limitations of this single value estimate of the convolved plasma properties along the line of sight.

%%%%%%%%%%%%%%%%%%%%%%%%%%%%%%%%%%%%%%%%%%%%%%%%%%%%%%%%%%%
\section{Results and Discussion}\label{sec:results}

To our knowledge, this is the first time that spatially resolved maps of the \ion{Si}{10} 1430 nm line characteristics have been obtained in the solar corona, either during an eclipse or from the ground.  Figure~\ref{fig:siX_maps}, in particular, demonstrates the results attainable from DKIST and Cryo-NIRSP, even when humidity conditions fluctuate, provided one carefully accounts for the impacts of telluric absorption during the background removal process.  The method introduced here (\S\ref{sec:background_removal}) bypasses the need for direct modeling of the entire telluric absorption profile, which, as in \citet{smette2015}, requires more detailed knowledge of the local atmospheric profile during the observations.  Instead, we model only the fluctuations in the water absorbance between the science observations and the flat-field calibration measurements using a single constant-property slab model.  By applying the flat field through division, we also help mitigate systematics in the instrument gain, such as interference fringes. The coronal line is fit with a Gaussian function; although, the method presented here could be extended to other model functions.  
 
For the observed active region, the 10th, 50th, and 90th percentile values for the peak line amplitude are 4.4, 8.4, and 15.8 $\mu$B$_{\odot}$, respectively. For the total line radiance, these values are 124, 224, and 412 photons cm$^{-2}$ arcsec$^{-2}$. \citet{penn1994_siX} reported a comparable value ($4.5~\mu$B$_{\odot}$) for their slit-averaged line center intensity within a small active region.  \citet{dima2019} inferred similar values from polarized brightness observations and modeled estimates of the linear polarized amplitude.  \citet{schad2020} forward synthesized \ion{Si}{10} line radiances through a 3D radiative magnetohydrodynamic simulation of the active corona and reported mean values of 34 photons cm$^{-2}$ arcsec$^{-2}$ using photospheric abundances; when scaled to coronal abundances, the value is 132 photons cm$^{-2}$ arcsec$^{-2}$, in reasonable agreement with these observations. 

The spatial loop structures in the \ion{Si}{10} maps (Figure~\ref{fig:siX_maps} correspond reasonably well with the EUV observations of similar characteristic temperature response, as shown in Figure~\ref{fig:sdo_dem} via a comparison of the AIA images with contour lines of the \ion{Si}{10} line peak intensity.  The 171 \AA\ and 193 \AA\ channels have a characteristic temperature near $10^{5.8}$ and $10^{6.2}$ in the non-flaring corona \citep{lemen2011}.  In each figure panel, the contour lines are given for the measured \ion{Si}{10} line peak amplitude for equal 2$\mu$B$_{\odot}$ samples between 5 and 19 $\mu$B$_{\odot}$.  Meanwhile, the reconstructed DEM map at $10^{6.15}$ K, corresponding to the equilibrium formation temperature of \ion{Si}{10} (see Figure~\ref{fig:sdo_dem} panel (c)), also shows a high degree of spatial correlation with the \ion{Si}{10} line intensity (given by the contours in all panels).  Some deviations are to be expected from the different time range of the slit-based raster observations and the 2 hour SDO/AIA average.  Furthermore, the spatial resolution of the SDO/AIA data is ~1.2'' is likely better in comparison to the seeing-limited conditions at DKIST on this date.

The median line center position of the \ion{Si}{10} coronal line is 1430.076 nm in these observations, which is in close agreement to that inferred by \citet{penn1994_siX}.  Our value is referenced to the telluric atlas spectrum in the NOAO/NSO FTS atlas and not adjusted for solar rotation or orbital motions.  90\% of the values lie within $\pm 1.6$ km sec$^{-1}$ of the median line position.  Some of the velocity patterns align with loop-like features in the field-of-view, as expected for slow flows along the coronal loops. 

The 10th, 50th, and 90th percentile values for the \ion{Si}{10} full width at half-maximum (FWHM) line widths are 0.245, 0.277, and 0.300 nm, respectively.  \citet{schad2023} reported median FWHM line widths of 0.27 $\pm$ 0.01 nm in a coronal streamer at elongations of 1.1 to 1.25 R$_{\odot}$, and these were in agreement with those reported by \citet{penn1994_siX} and \citet{dima2019}.

In section~\ref{sec:densityMethod} we outlined a method for estimating the electron density along the line-of-sight by combining the EUV-reconstructed DEM and the measured line radiances of the \ion{Si}{10} line whose emissivity (\textit{i.e.}, contribution function) is density sensitive.  The results shown in panel (d) of Figure~\ref{fig:sdo_dem} indicate electron densities in the range of $10^{9}$ to $10^{8}$ cm$^{-3}$ decreasing with height, which are compatible with electron densities previously reported at the range of heights observed here \citep[see,\textit{e.g.},][]{mason1999,madsen2019}.  It is also noted that the active region core is already slightly rotated onto the observable solar disk, as shown in panels (a) and (b) of the figure, and so these observations sample higher altitudes of the active region plasma.  Of course, a number of uncertainty sources exist that may affect these results, including those inherent to the DEM inversion processes, elemental abundances, the absolute photometric accuracy of the Cryo-NIRSP observations, and non-uniform densities along the line-of-sight.  In the future, the method introduced here can be evaluated in comparison to more direct line-ratio methods, especially of the \ion{Fe}{13} lines at 1074 and 1079 nm.

The excess width of coronal emission lines, beyond that induced by thermal broadening, is well known \citep[see, \textit{e.g.},][among others]{boland1973, doschek1976, singh2006, tomczyk2007} and may be used to investigate the role of wave motions and/or turbulence in the coronal energy balance \citep[see, \textit{e.g.},][]{mcIntosh2012,fyfe2021} The non-thermal velocities derived in \S\ref{sec:nonthermal} and shown in Figure~\ref{fig:nonthermal} range from 8 to 23 km sec$^{-1}$, largely consistent with the previous measurements of nonthermal line widths and larger than the CryoNIRSP instrumental line width ($\approx c/45000 \simeq 6.5$ km sec$^{-1}$).  While the measurement errors are expected to depend on the signal strength, we can estimate the error in the derived nonthermal velocity for the spectrum repeatedly fitted in Appendix~\ref{sec:appendix1} (see Figure~\ref{fig:corner}).  Here, the measured FWHM and its error is 0.265 nm ( -7 pm / + 11 pm) for 25\% changes in the optimal objective function value.  For a temperature of 10$^{6.15}$ K, the nonthermal line width is 16.2  km sec$^{-1}$ with estimated errors of $-1.9$ km sec$^{-1}$ and $+2.7$ km sec$^{-1}$. 

Using HINODE/EIS data, \citet{hara2008} and \citet{hahn2012} have studied non-thermal line widths in active regions and polar coronal holes, respectively.  The former show enhanced non-thermal velocities near the footpoints of active regions, when observed on-disk.  The latter show an increase and subsequent decrease in non-thermal line widths with higher projected heights above the solar limb, and with amplitudes slightly larger than our active region measurements.  In the Cryo-NIRSP observations, we see the lowest non-thermal line widths (${\sim}15$ km sec$^{-1}$) in the brightest regions of the active region (see histogram in panel (b) of Figure~\ref{fig:nonthermal}), near the lower portions of the loops in the center of the field-of-view.  The region near $\left<X,Y\right> = \left<50,1050\right>$ shows comparative larger line-widths ($\gtrsim 20$ km sec$^{-1}$).  We note that this region is hotter according to the $T_{eff}$ shown in Figure~\ref{fig:sdo_dem} panel (f).  It is also a region of increased temporal dynamics in AIA 193\AA\, which leads to larger relative errors in the DEM-reconstructed intensities shown in panel (d). Similar to \citet{hahn2012}, we do see increased nonthermal line widths in the quiescent area above the active region near  $\left<X,Y\right> = \left<0,1130\right>$; though, these areas also have comparatively weaker line amplitudes (${\sim}$5 to 7$\mu$B$_{\odot}$).  It is possible that this is a physical effect; however, we cannot at this time rule out systematic biases introduced by the blended telluric absorption, especially when the line signal is weak in comparison to the magnitude of the scattered light.  In these observations, the scattered light magnitude ranges from ${\sim}250~\mu$B$_{\odot}$ near the limb to ${\sim}80~\mu$B$_{\odot}$ at the outer edge of the field of view.  We expect future observations will improve upon these commissioning data for a more systematic (and multi-wavelength) study of nonthermal line widths.

%%%%%%%%%%%%%%%%%%%%%%%%%%%%%%%%%%%%%%%%%%%%%%%%%%%%%%%%%%%
\section{Summary and Outlook}\label{sec:discussion}

We have presented commissioning observations of the \ion{Si}{10} 1430 nm forbidden coronal emission line by DKIST/Cryo-NIRSP.  These results demonstrate new capabilities for studying the \ion{Si}{10} forbidden coronal emission line at 1430 nm, which potentially provides valuable polarimetric diagnostics for the coronal magnetic field as well as additional diagnostics for nonthermal velocities.  Despite challenges posed by telluric absorption, this work has successfully extracted the line characteristic across a scan of coronal active region, even in unstable observing conditions.  We have proposed a method for treating fluctuations in the amount of telluric water content that allow better extraction of the line signal from its scattered light background.  In future observations, it is recommended that multiple flat-field observations at different airmasses be obtained to assist in the separation of the telluric absorption and the solar spectrum, and so that methods may be further refined.  We have further validated these observations, and the existing knowledge of the formation of the \ion{Si}{10} 1430 nm, by comparing the observations to EUV observations and deriving realistic and self-consistent estimates for the electron density, effective temperature, and non-thermal line widths in the corona. These advances represent a significant step forward for observations of this line, and lend further support for advancing polarized observations in this bandpass in pursuit of magnetic field diagnostics.  Well-calibrated polarimetric observations may also aide in the separation of the multiple solar and telluric signals that contribute to the observed flux due to the unique conditions that polarize these components \citep[see, \textit{e.g.},][]{schad2022}.
\\
\\
\noindent
The research reported herein is based in part on data collected with the Daniel K. Inouye Solar Telescope (DKIST), a facility of the National Solar Observatory (NSO). NSO is managed by the Association of Universities for Research in Astronomy, Inc., and is funded by the National Science Foundation. Any opinions, findings and conclusions or recommendations expressed in this publication are those of the author(s) and do not necessarily reflect the views of the National Science Foundation or the Association of Universities for Research in Astronomy, Inc.  DKIST is located on land of spiritual and cultural significance to Native Hawaiian people. The use of this important site to further scientific knowledge is done so with appreciation and respect.  The SDO data are provided courtesy of NASA/SDO and the AIA science team. G. Dima was supported in this research by NOAA cooperative agreements NA17OAR4320101 and NA22OAR4320151.

\facility{DKIST}

\appendix
\section{Model degeneracies}\label{sec:appendix1}

\begin{figure*}
\centering
\includegraphics[width=0.93\textwidth,clip]{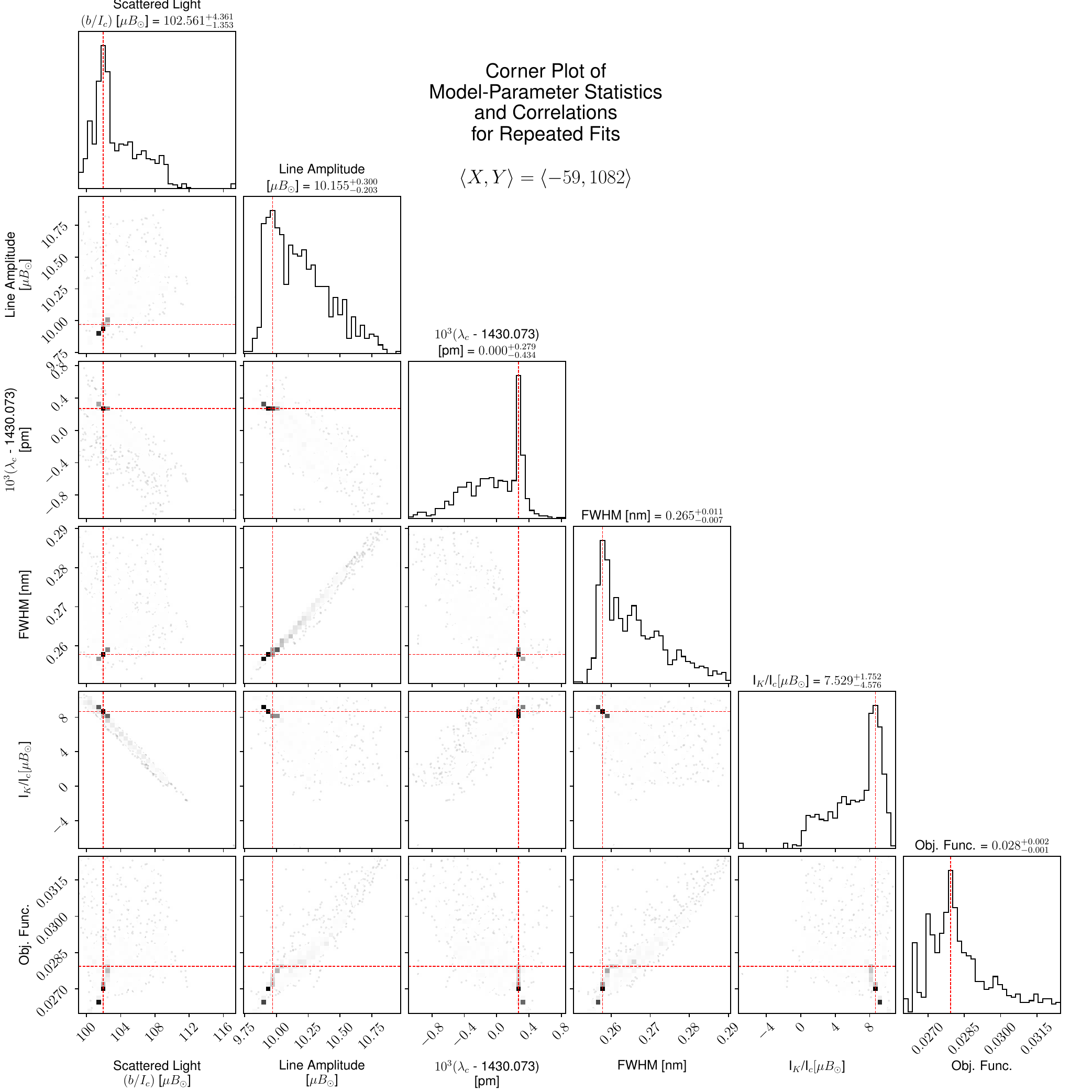}
\caption{Corner plot of model-parameter statistics and correlations for 900 model fits to the spectrum observed at $\left <X,Y \right> = \left < -59, 1082 \right>$. To improve plot readability, only the parameters relevant to the coronal line values are plotted.  The top of each column is 1d histogram of the parameter listed at the bottom, while all other panels are 2d histograms. $\lambda_{c}$ in the third row and column (from top and left) refers to the fitted line center wavelength of the coronal line. Here we show its deviation from the best fit in units of picometers.
\vspace{-5mm}
}
\label{fig:corner}
\end{figure*}

As discussed in Section~\ref{sec:objective}, our model is intended to extract the \ion{Si}{10} coronal line parameters while the other parameters are allowed to contain degenerate information.  We investigated parameter correlations by performing 1000 repeated optimizations for the spectrum at $ \left <X,Y \right> = \left < -59'', 1082'' \right> $, as also plotted in right panels of Figure~\ref{fig:fitting_examples}.  The first step in the optimization involves the Differential Evolution method with a randomized initial population, which ensures stochastic realization of the optimized model parameters.  The correlation analysis is limited to the 900 optimized fits with a objective function value within 25\% of the minimum value of all fits.  We find strong correlations, as expected, for the telluric slab model parameters ({i.e.}, between the mixing ratio and total slab thickness); however, we do not find concerning correlations for the coronal line parameters.  The corner plot \citep{corner} of Figure~\ref{fig:corner} shows the 1d and 2d histograms of parameters relevant to the 3 values describing the Gaussian coronal line.  The errors implied by the width of the 1d histograms is relatively small for the coronal line parameters.  Finally, we include panels for the scattered light magnitude and the K-corona intensity to demonstrate that these parameters are degenerate with each other, which is expected given the dominance of telluric absorption over the strength of the scattered photospheric lines.

\vspace{-2mm}

\bibliography{main}{}
\bibliographystyle{aasjournal}

\end{document}